# Textual Spatial Cosine Similarity

Giancarlo Crocetti
*Pace University Seidenberg School of CSIS, White Plains, NY 10606, USA*
*gcrocetti@pace.edu*

*Abstract*— When dealing with document similarity many methods exist today, like cosine similarity. More complex methods are also available based on the semantic analysis of textual information, which are computationally expensive and rarely used in the real time feeding of content as in enterprise-wide search environments. To address these real-time constraints, we developed a new measure of document similarity called Textual Spatial Cosine Similarity, which is able to detect similitude at the semantic level using word placement information contained in the document. We will see in this paper that two degenerate cases exist for this model, which coincide with Cosine Similarity on one side and with a paraphrasing detection model to the other.

## I. INTRODUCTION

From a theoretical perspective similarity can be defined as a quantifiable measure of how similar two objects are, which can be applied to textual information as well.

Many measures of similarity between documents exists today [2][5][9], and cosine similarity is widely used in retrieval systems [5][9] today. All search engines available today, both commercially and open sourced, provide the implementation of cosine similarity when comparing documents indexed by the system.

Despite its popularity, the cosine similarity has the drawback of not considering word placement in the text under analysis. A classic example of this problem is the comparison of the two texts "John loves Mary" and "Mary loves John". These two simple sentences use the same words, but their meaning is completely different. When comparing these documents, the cosine similarity will produce the exact-match value of 1, due to identical term vectors, yet this result is arguable to say the least.

Methods based on Natural Language Processing (NLP) do exists that consider the semantic context within each document, but such methods are computationally more intensive when compared to cosine similarity due to the multitude of algorithms applied to the text, which increase, considerably, the associated computation overhead [1].

In this paper we introduce a textual space similarity measure, which provides semantic-level quality, but without the overhead of semantic approaches that might not scale up when processing queues of millions of documents. In all examples in this paper we consider documents in which we applied tokenization, stop-wording and stemming processes.

## II. TEXTUAL SPACE SIMILARITY

To overcome this lack of spatial analysis in the cosine similarity, we introduce a new measure of similarity: Textual Space Similarity. This approach is different from methods based on positional language models [7][8] focused on improving ranking results based on the matching locality of the query terms in documents. It is also different from the approach used in tasks requiring the computation of the similarity between two very short segments of text [6].

Let $d_i$ and $d_j$ be two documents and let $p_i^{t^k}$ and $p_j^{t^k}$ be the ordinal position of the *k*-instance of the term *t* (starting from 0) in the document $d_i$ and $d_j$ respectively when the *k*-instance of the term *t* appears in both documents, 0 otherwise.

We define the *Spatial Difference* for the term *t* as the quantity $sd_{i,j}^t = \sum_k \frac{\left|p_i^{t^k} - p_j^{t^k}\right|}{p_i^{t^k} + p_j^{t^k}}$.

Each term in the summation can be 0 when the *k*-instance of the term *t* appears in exactly the same ordinal position in both documents and 1 when *t* is in the initial position of 0 in $d_i$ and in some other position $p_j^t \neq 0$ in $d_j$ (or vice versa). Therefore the quantity $\frac{\left|p_i^{t^k} - p_j^{t^k}\right|}{p_i^{t^k} + p_j^{t^k}} \in [0,1]$.

Finally we define the Textual Space Similarity of the two documents $d_i$ and $d_j$ to be the quantity:

$$TSS(d_i, d_j) = \frac{\sum_t sd_{i,j}^t}{\lambda}$$

In the numerator we have the sum of the Spatial Differences $sd_{i,j}^t$ between documents $d_i$ and $d_j$ for all common terms *t*, and at denominator we have the number of matches $\lambda$.

Because the numerator is the summation of quantities between [0,1] appearing no more than $\lambda$ times, the largest value that $TSS(d_i, d_j)$ can assume is $\frac{1+1+\cdots+1}{\lambda \; times} = \frac{\lambda}{\lambda} = 1$ and the smallest value of 0 when two documents are exactly the same. Therefore $0 \leq TSS(d_i, d_j) \leq 1 \; \forall \; d_i, d_j$.

In order for $TSS(d_i, d_j)$ to have the same direction as other measures of similarity, 1 for an exact match and 0 for total dissimilarity, we modified its definition in its final form:

$$TSS(d_i, d_j) = 1 - \frac{\sum_t sd_{i,j}^t}{\lambda}$$

If we consider the "John loves Mary" and "Mary loves John" example we have:

$$TSS(\text{"John loves Mary"}, \text{"Mary loves John"}) =$$
$$= 1 - \frac{\sum_t sd_{i,j}^t}{\lambda} = 1 - \frac{\frac{|0-2|}{2} + \frac{|1-1|}{2} + \frac{|2-0|}{2}}{3} =$$
$$= 1 - \frac{2}{3} = 0.33$$

This result is quite different from the exact match produced by the cosine similarity and better captures the spatial and semantic differences between the two texts.

### III. TEXTUAL SPATIAL COSINE SIMILARITY

In order to consider both spatial and word features in documents we can combine the Cosine Similarity (*sim*) and the Textual Space Similarity (*TSS*) using a weighted approach and finally define the Textual Spatial Cosine Similarity between documents $d_i$ and $d_j$ as:

$$TSCS(d_i, d_j) = \alpha * sim(d_i, d_j) + (1 - \alpha) * TSS(d_i, d_j)$$

with $\alpha \in [0,1]$.

A value of $\alpha=0$ will make TSCS coincide with the textual space similarity $TSCS(d_i, d_j) = TSS(d_i, d_j)$ and a value of $\alpha=1$ will make TSCS coincide with the cosine similarity $TSCS(d_i, d_j) = sim(d_i, d_j)$.

Deciding which value to assign to $\alpha$ will depend to the problem at hand, but for quality textual documents this can safely be set at 0.5. In other situations, especially with automatic extraction of text from web articles, the spatial information might not be accurate due to the presence of extraneous characters and therefore an higher value for $\alpha$ might be warranted.

Considering, once again, the previous example we have:

$$TSCS(\text{"John loves Mary"}, \text{"Mary loves John"}) =$$
$$= 0.5 * 1 + 0.5 * 0.33 = 0.67.$$

This value indicates a better balance between the undeniable similarity of the texts, their semantic, and spatial differences when compared to the cosine "exact match".

### IV. SIMILARITY THRESHOLD

Since $TSCS(d_i, d_j)$, as for the cosine similarity, still depends on the size of the corpus and on the frequency of terms across documents, it is not a trivial matter to identify a similarity threshold that will tell us when two documents are similar. In fact, the document frequencies will tend to decrease as the size and number of documents in the corpus increase, since the probability of the same term appearing on different documents will increase as well.

In order to have a sense of how the textual spatial cosine similarity value for two sets of seeded documents varies with corpora of different sizes, we setup an experiment in which the similarity of two sets of seeded documents is measured as we add random documents to the corpora. In this experiment we used a value of $\alpha=0.5$.

The first set of seeds is comprised of two documents represented by the same article in which we changed very few terms. This pair has a very high similarity value. The second set of seeds is represented by two documents taken from two different newspapers about a related subject. Such documents would be considered similar to a human reader. The results are shown in Tables 1 (a) and (b) in which we also compared the influence of the corpus size for both measures of similarity: Cosine and TSCS.

The experiment was repeated several times with analogous results.

The TSCS showed a lesser influence to the corpus size with smaller variations in the results ranging from 0.03 to 0.07. Comparing this to the Cosine function, that revealed variations as large as 0.12, we can safely leave the similarity threshold at the default value of $\alpha=0.5$.

Table 1 (a) – Similarity variations with different corpus sizes using TSCS.

| Size of Corpus | Similarity of Set #1 | Similarity of Set #2 |
|---|---|---|
| 4 | 0.89 | 0.52 |
| 5 | 0.89 | 0.53 |
| 10 | 0.90 | 0.54 |
| 15 | 0.90 | 0.54 |
| 20 | 0.91 | 0.56 |
| 30 | 0.92 | 0.57 |
| 40 | 0.92 | 0.59 |

Table 1 (b) – Similarity variations with different corpus sizes using Cosine similarity.

| Size of Corpus | Similarity of Set #1 | Similarity of Set #2 |
|---|---|---|
| 4 | 0.85 | 0.48 |
| 5 | 0.85 | 0.50 |
| 10 | 0.87 | 0.51 |
| 15 | 0.86 | 0.52 |
| 20 | 0.89 | 0.44 |
| 30 | 0.90 | 0.57 |
| 40 | 0.91 | 0.60 |

### V. TSCS IN DETECTING PARAPHRASING

As final test we considered the classic example of paraphrasing [3][4], often used to highlight the shortcomings of cosine similarity as a way of introducing alternative approaches based on semantic models.

Let's consider the following texts:

Text 1: "When the defendant and his lawyer walked into the court, some of the victim supporters turned their backs to him."

Text 2: "When the defendant walked into the courthouse with his attorney, the crowd turned their backs on him".

The problem is whether to consider these two phrases as paraphrase or not. We will consider two documents a paraphrase if their similarity measure is 0.5 or above.

The computation of cosine similarity produced the matrix shown in table 2a.

**Table 2 (a)** – Cosine Similarity on the paraphrase problem

|        | Text 1 | Text 2 |
|--------|--------|--------|
| Text 1 | 1.0    | 0.31   |
| Text 2 |        | 1.0    |

**Table 2 (b)** – TSCS on the paraphrase problem

|        | Text 1 | Text 2 |
|--------|--------|--------|
| Text 1 | 1.0    | 0.60   |
| Text 2 |        | 1.0    |

The cosine similarity value of 0.31 clearly misses to consider these documents as paraphrase, however the TSCS measure of 0.60 (table 2b) is able to capture the semantic resemblance of the texts by clearly categorizing these two texts as paraphrase.

We extended our experiment to the paraphrase dataset located at http://www.cs.york.ac.uk/semeval-2012/task6/data/uploads/dataset/train.tgz, and used in the Semantic Textual Similarity Task from the University of York in UK, which contains paraphrases of variable lengths.

The dataset consisted of 734 English pairs drawn from publicly available datasets:
1. Microsoft Research Paraphrase Corpus
2. Microsoft Research Video Description Corpus
3. WMT2008 development dataset

We analyzed the TSCS performance in detecting paraphrases, by using different values of alpha as reported in figure 1.

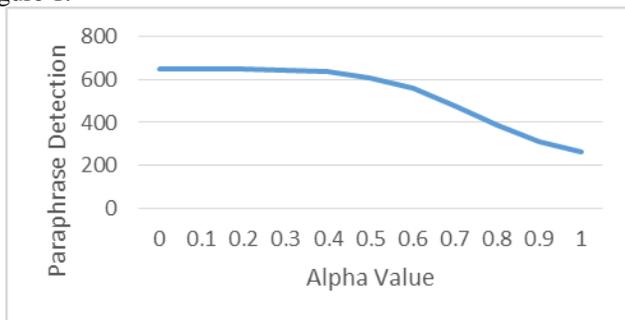

**Figure 1** – TSCS performance on the paraphrase dataset

As shown in the diagram, we detected the maximum number of paraphrases for α=0, in which the TSCS recognized a total of 649 paraphrases with an accuracy of $\frac{649}{734} = 0.8842$, which is quite high. Consequently, TSCS in its degenerate case of TSS (α=0), can be successfully adopted in the detection of paraphrases with high level of accuracy when compared to the cosine similarity that detected 261 cases representing only 36% of the total.

VI. CONCLUSIONS AND FUTURE WORK

The Textual Space Cosine Similarity (TSCS) adds a spatial dimension to the problem of document similarity without the need of more complex, and computational intensive, semantic approaches.

Through experimentation we shown that TSCS is minimally sensitive to changes in the corpus size, and that not only TSCS is able to improve cosine similarity in considering semantic differences in texts, but it can be used as a model for paraphrasing detection with accuracy levels close to 90%. This outcome has been backed up with extensive tests on a large dataset of paraphrases.

By varying the parameter alpha in the TSCS formulation we are able to tune the model to fit specific needs within the two degenerative case:
1. α=0: coincide with Textual Space Similarity and proved to be a good model for paraphrasing detection.
2. α=1: coincide with Cosine Similarity.

TSCS can be used by search engines to improve document similarity by replacing the traditional cosine similarity approach or in other areas as in the following examples:

1. Detection of plagiarism
2. Content recommendation
3. Content Discovery

As future work, the results of this study can be strengthened by comparing TSCS with other similarity measures beyond cosine, and utilizing datasets widely used for evaluating feature selection techniques, classification and clustering.